\documentclass[aps,prd,preprint,floatfix]{revtex4}
\usepackage{epsfig}
\usepackage{graphicx}
\usepackage{amsmath,amssymb,amsfonts,latexsym}
\usepackage{epsfig,amssymb,amsfonts,verbatim}
\usepackage{amsmath}
\usepackage{latexsym}
\usepackage{amsfonts}
\usepackage{amssymb}

\begin{document}

\title{Time-reversal operation and transport phenomena in topological insulators}
\author{Andrew Das Arulsamy}
\affiliation{Condensed Matter Group, Institute of Interdisciplinary Science, No.~24, level-4, Block C, Lorong Bahagia, Pandamaran, 42000 Port Klang, Selangor DE, Malaysia}

\maketitle
~\\
\textbf{Abstract:} We prove that any forms of electric current measurements within a gapped or gapless quantum system necessarily violate the time reversal symmetry (TRS). We then use the ionization energy theory to unequivocally show that the metallic surface states of a topological insulator consist of some `special' energy level crossings, not due to TRS induced Kramers degeneracy because there is a finite energy gap due to different wavefunctions. We use this special crossings and derive the electron-ion scattering rate required to explain the resistivity and carrier-type transition in (Bi$_{1-x}$Sb$_x$)$_2$Te$_3$ and Pb$_{1-x}$Sn$_x$Se topological insulators.
$^*$Corresponding author, E-mail: sadwerdna@gmail.com
\newpage

\section*{1. Introduction}

Degenerate energy levels (gapless) can form conducting states even if they are confined to the surface (or to the edge). If their degeneracy is kept intact against external disturbances (temperature ($T$), pressure, electric ($\textbf{E}$) and magnetic ($\textbf{B}$) fields), such that the bulk always has a well-defined band gap, then the materials satisfying the above conditions can be grouped into a new class of solids. These new solids are known as the topological insulators (TI). At least, that is the proper and generalized definition one can think of to identify TI based on the crossed energy-level notion in two-dimensional quantum Hall metals, which was first calculated by Hatsugai~\cite{hat}. This definition also applicable to three-dimensional TI~\cite{mele,balent}, and it properly rules out free-electron and Fermi-liquid metals, even in the presence of skin effect (frequency-dependent surface conductivity)~\cite{ash}. In TI, we actually need the bulk to be gapped so that the surface states form a distinct two-dimensional system that can exhibit quantum Hall metallic properties~\cite{rui,chang}. This means that, one may need to perform quantum Hall effect measurements~\cite{rui,chang,konig} to confirm whether an insulator is indeed a topological insulator because electric transport measurements~\cite{barua,vonk}, and other surface analysis techniques, such as the angle-resolved photoemission spectroscopy (ARPES)~\cite{ylchen}, de Haas-van Alphen~\cite{kittel} and Shubnikov-de Haas~\cite{shub} effects are insufficient to claim the bulk is gapped, while the surface is metallic, independent of the bulk.   

On the other hand, TI are also related to topology, a mathematical notion that deals with `smooth' deformations of any entity~\cite{naka}, and in our case, this entity strictly refers to two-dimensional surface metallic states or degenerate energy levels. One can think of these metallic states or energy levels as the elements (or open subsets) within an open set that comprises all degenerate energy levels. This means that the energy levels or the subsets are the topology on an open set where this open set and its subsets satisfy some precise conditions, which allow smooth deformations~\cite{geroch}. Now, any deformation to these energy levels requires one to invoke an arbitrary right-hand side (rhs) action operator~\cite{naka} that tacks a transition function (namely, a phase factor) on the rhs of an electronic wavefunction~\cite{and1}. However, it has been rediscovered recently that a wavefunction picks up or drops a phase factor as a result of a physical notion known as the Pancharatnam phase retardation (due to a phase acceleration or deceleration). In fact, when the phase and(or) group momenta(um) of a particular wavefunction changes, then one has to tack the Pancharatnam-Berry (PB) phase factor on the rhs of a wavefunction, which gives rise to the Pancharatnam wavefunction transformation~\cite{and2} that was originally discovered (formulated and observed) by Pancharatnam using the polarized pencil beams and Poincar${\rm \acute{e}}$ sphere~\cite{pan1,pan2}. This means that, the original Berry's phase~\cite{berry} is a rediscovered special case (the group momentum of a wavefunction is invariant) within the generalized Pancharatnam phase~\cite{and1,and2,pan1,pan2}. Compared to a conducting surface, the bulk requires the wavefunction to be transformed beyond the phase factor due to a nonzero energy gap between the valence and conduction bands. Therefore, the above rhs action operator has a precise physical origin within the context of quantum mechanics due to the Pancharatnam wavefunction transformation~\cite{and1}. 

Having said that, we can now set the next course of action---to understand how the definition for a topological insulator (stated in the first paragraph) can be achieved physically, or, how to identify the relevant physical parameter(s) responsible to obtain such an insulator. In this respect, Kane and Mele~\cite{kane}, and independently by Bernevig, Hughes and Zhang~\cite{bern2} have confirmed the Hatsugai's surface energy-level crossings such that one needs a combination of physical properties to activate the required energy-level crossings at the edge or on the surface, while the bulk is of course, remains gapped. The physical phenomena that are at play to produce both gapped bulk and gapless surface states in a single system come from different magnitudes of certain crucial parameters (the bulk has different magnitudes compared to the edge or surface)~\cite{kane}. For example, the interplay among these critical parameters, namely, the spin-orbit coupling, intersite hopping matrix elements, Rashba coupling and the staggered sublattice potential gives rise to a topologically insulating phase~\cite{kane}. 

Note this, the above critical parameters that eventually give rise to changing electronic structure in the bulk can only be different from its surface if their respective chemical composition and/or the coordination numbers are themselves different. For example, HgTe/Hg$_{0.3}$Cd$_{0.7}$Te quantum well for different thickness ($d$) have given rise to a quantum phase transition (QPT) from the usual gapped surface (similar to bulk) to a gapless surface states (while the bulk is still gapped) where the critical thickness here is $d_{\rm c}$ = 6.3 nm that refers to the width of their devise. However, the existence of odd numbers of spin-polarized edge channels is actually an assumption~\cite{konig}. But never mind, the said QPT has been correctly associated to changing electronic structure due to spin-orbit coupling (SOC)~\cite{konig} following the Bernevig-Hughes-Zhang (BHZ) model~\cite{bern2}. Here, it is alright to not to know exactly what other critical parameter(s) (besides SOC) are responsible for the changing electronic structure (or changing energy(Landau)-level crossing). Anyway, what we wanted to say here is that the relation between changing Landau-level crossing and $d$ due to SOC originate from the changing defect types and composition or more precisely, from changing chemical composition and/or the coordination numbers with changing $d$. These changes due to defect types and composition will give rise to changing SOC~\cite{arul9}, which shall be made explicit when we discuss the doping-dependent resistivity in TI.
   
The point that we do not agree with K${\rm \ddot{o}}$nig \textit{et al}.~\cite{konig} and BHZ~\cite{bern2} is the assumption that time reversal symmetry (TRS) is not violated for the above devise for different applied gate voltages on the basis of BHZ model~\cite{bern2}. This invariance of TRS in the presence of electric current (due to spin or charge or both) in any condensed matter system should not be assumed to be true haphazardly due to a well thought-out hypothesis put forth by Messiah~\cite{albert}. See the additional notes (prior to conclusions) for proper arguments. In particular, the TRS defined in Ref.~\cite{albert} leading to Kramers degeneracy~\cite{kramers} is not necessarily required to obtain crossed energy levels due to interplay among the above stated critical physical parameters. Of course, one can construct a particular Hamiltonian to satisfy TRS if the energy levels are crossed (not necessarily Dirac points due to Kramers degeneracy) between the bulk valence and conduction bands~\cite{konig,ylchen,hasan} such that $E_1(\textbf{k}) = E_2(-\textbf{k})$ where $\textbf{k}$ denotes the wavevector~\cite{rroy3}. But this construction does not imply that TRS is always preserved by default in the presence of internal electric current---for example, all standard ($\textbf{B} = 0$) and Hall resistance ($\textbf{B} \neq 0$) measurements~\cite{hall1,klit,tsui} necessarily violate TRS as a result of nonzero internal electric current (static or time-dependent)~\cite{albert}. For Hall current, one needs to consider Zeeman splitting~\cite{zeeman1} due to nonzero applied magnetic field, while static current produces static `external' magnetic ($\textbf{B}'$) field. In our formalism within IET, we will properly consider TRS to address electron conduction in TI.

Therefore, our objective here is to develop the basic mechanism required to consistently explain the electronic transport phenomena and carrier-type transition in TI down to atomic energy levels. We start our analyses by invoking the properties of ionization energies and energy level spacings within the ionization energy theory (IET). Subsequently, we derive the relevant equations required to properly understand the time reversal operation in TI, and then justify the conditions that may give rise to TRS violation in TI. With these knowledge as background, we will first evaluate the influence of the energy-level spacing ($\xi$, which is also known as the ionization energy) on resistivity and carrier-type transition with respect to different chemical compositions in (Bi$_{1-x}$Sb$_x$)$_2$Te$_3$ and Pb$_{1-x}$Sn$_x$Se TI. The band structure properties of (Bi$_{1-x}$Sb$_x$)$_2$Te$_3$ and Pb$_{1-x}$Sn$_x$Se TI relevant to understand transport phenomena have been investigated by Jinsong Zhang \textit{et al.}~\cite{jins} and Dziawa \textit{et al.}~\cite{dzia}, respectively. 

We will exploit the relation between $\xi$ and the electron excitation probability to answer the microscopic origins for (i) the changes in doping-dependent resistivity data (including the carrier-density and electron-ion scattering magnitudes) in (Bi$_{1-x}$Sb$_x$)$_2$Te$_3$ and Pb$_{1-x}$Sn$_x$Se TI, and (ii) the carrier-type transition ($\texttt{p}$- to $\texttt{n}$-type or vice versa) for different chemical compositions and defects in (Bi$_{1-x}$Sb$_x$)$_2$Te$_3$. Our research on the above-stated points ((i) and (ii)) can lead us to understand how TI fit properly and consistently within the current knowledge of atoms, metallic Fermi liquid, semiconductors, ferromagnets, Mott-Hubbard insulators, cuprates strange metals and quantum Hall metals. In addition, these points were not addressed nor generalized in earlier reports with respect to different types of atoms, and with respect to non-Kramers degeneracy (crossed energy levels with $\xi \neq 0$). 

\section*{2. Ionization energy theory}

Ionization energy is defined as the minimum energy required to remove a bound electron (bounded to a nucleus of an atom) to infinity ($\textbf{r} \rightarrow \infty$). In contrast, IET is based on the theorem that states---the minimum energy required to excite or to polarize a bound electron to a finite distance ($\textbf{r} \rightarrow \textbf{r}_{\rm finite}$) within a quantum system (atomic, molecular or solid) is proportional to the ionization energy defined above. This proportionality is precisely known as the ionization energy approximation~\cite{arul1,arul2}, which can be written as,
\begin {eqnarray}
\xi^{\rm quantum}_{\rm system} \propto \xi^{\rm constituent}_{\rm atom}. \label{eq:1}
\end {eqnarray}  
Within a quantum system, the energy is quantized and therefore, the ionization energy is also known as the energy level spacing. This energy level spacing can be used to rewrite the standard Schr${\rm \ddot{o}}$dinger equation to read~\cite{arul3,arul4},
\begin {eqnarray}
&&{\rm i}\hbar\frac{\partial \Psi(\textbf{r},t)}{\partial t} = \bigg[-\frac{\hbar^2}{2m}\nabla^2 + V_{\rm IET}\bigg]\Psi(\textbf{r},t) \nonumber \\&& = H_{\rm IET}\Psi(\textbf{r},t) = (E_0 \pm \xi)\Psi(\textbf{r},t), \label{eq:2}
\end {eqnarray}  
which is known as the IET-Schr${\rm \ddot{o}}$dinger equation where $H_{\rm IET}$ and $V_{\rm IET}$ are the exact Hamiltonian and potential term, respectively. The properties of electrons are represented by the true and real (not a guessed) wavefunction, $\Psi(\textbf{r},t)$, $\hbar$ is the Planck constant divided by 2$\pi$ and $m$ denotes electron mass. Here $E_0$ is the ground state energy for temperature ($T$) equals zero Kelvin, in the absence of external disturbances. The sign, $\pm$ refers to electrons and holes, respectively. The above generalized definitions for $H_{\rm IET}$ and $V_{\rm IET}$ imply Eq.~(\ref{eq:2}) is too general to capture a specific quantum matter. One way to solve Eq.~(\ref{eq:2}) is to rewrite it in the usual form, 
\begin {eqnarray}
\bigg[-\frac{\hbar^2}{2m}\nabla^2 + V\bigg]\Psi(\textbf{r},t) = H\Psi(\textbf{r},t) = E\Psi(\textbf{r},t), \label{eq:3}
\end {eqnarray}  
after which, one can incorporate the relevant potential terms ($V$) with appropriate approximations, and make do with the guessed wavefunction, satisfying the variational principle to find the energy eigenvalues. Here, we will not follow this approach. Instead, we make use of Eqs.~(\ref{eq:1}) and~(\ref{eq:2}) to tackle a specific quantum matter. For example, $\xi$ will furnish one with the details on the electron excitation probabilities~\cite{arul3} for a given material with different chemical compositions~\cite{arul1,arul2}. These probabilities can be properly incorporated into the relevant physical parameters (namely, carrier density, electron-ion interaction term and carrier-type transition) by means of the energy-level spacing renormalization group method~\cite{arul4}. Our renormalization procedure will renormalize the above physical parameters such that they can be used to determine the microscopic mechanisms responsible for doping-dependent electrical resistivity and carrier-type transition in TI. A specific potential term in Eq.~(\ref{eq:2}) has been exploited earlier in Refs.~\cite{qpt,ion}. Note this, one can also reprocess the energy-level spacing cut-off parameter exactly within the Shankar's wavenumber-dependent renormalization technique~\cite{shank,shank2,shank3}. For an obvious reason (see below), we prefer to work with the energy-level spacing. 

\subsubsection*{2.1. Ionization energies for isolated atoms}

We now list the chemical elements needed to make TI, calculate their average ionization energies and explain what these averaged values mean. For an hypothetical atom, $\texttt{X}^{i+}$ (with $i$ number of valence electrons) the averaged $\xi$ can be calculated from,
\begin {eqnarray}
\xi^{\rm constituent}_{\rm atoms} = \sum_j\sum_i^z \frac{1}{z}\xi_{j,i}(\texttt{X}^{i+}_{j}). \label{eq:4}   
\end {eqnarray}
Here, the subscript $j$ identifies the different types of chemical elements ($\texttt{X}_j$) in a given TI, while the subscript, $i = 1, 2, \cdots, z$, counts the number of outer electrons required to interact with its nearest and next nearest neighbors by forming bonds. We list all the essential chemical elements required to form TI in Table~1. Before averaging, the experimental ionization energy values for each outer electron (1$^{\rm st}$, 2$^{\rm nd}$, and so on) were obtained from Refs.~\cite{web,web2}. This is the only external information one needs to proceed. Table~1 also lists the averaged ionization energy values calculated from Eq.~(\ref{eq:4}) with respect to the stated valence states.    
We now briefly state the importance of approximating $\xi^{\rm quantum}_{\rm matter}$ from $\xi^{\rm constituent}_{\rm atom}$ ($\xi$ for short) via Eqs.~(\ref{eq:1}) and~(\ref{eq:4}). A chemical element, $\texttt{X}_j$ with large $\xi(\texttt{X}_j)$ means its valence electrons are not easily excited or polarized, and conversely, the electron-excitation probability and the atomic polarizability of a chemical element ($\texttt{X}_{j+1}$) are large if its $\xi(\texttt{X}_{j+1})$ is small. Furthermore, one can readily use Eq.~(\ref{eq:1}) to claim that the above probability and polarizability for a given TI get smaller if we systematically substitute one of its constituent chemical elements ($\texttt{X}_j$, $\xi(\texttt{X}_j)$) with another atom ($\texttt{X}_{j+1}$) that has a larger $\xi(\texttt{X}_{j+1})$. The above probability can be derived from the Fermi-Dirac statistics (FDS) with $\xi$ as an additional restrictive condition~\cite{arul1,arul2}, and is given by,
\begin{eqnarray}
f(E_0,\xi) = \frac{1}{e^{\lambda[(E_0 + \xi) - E_{\rm F}^{0}]}+1}, \label{eq:5}
\end{eqnarray} 
where $\lambda = (12\pi\epsilon_0/e^2)a_{\rm B}$, $\epsilon_0$ is the permittivity of free space, $a_{\rm B}$ is the Bohr radius and $E_{\rm F}^{0}$ denotes the Fermi level for $T = 0$K and without any external disturbances. Equation~(\ref{eq:5}) is also known as the ionization energy based FDS ($i$FDS). The atomic polarizability (due to outer electron displacement) can be determined from~\cite{arul5,arul6},
\begin {eqnarray}
\alpha(\xi) = \sum_j\sum_i\frac{Z_je^2}{m}\exp\big[\lambda(E_{\rm F}^0 - \xi_i)\big]\frac{f_i}{(\omega_{0i}^2 - \omega^2)}, \label{eq:6}
\end {eqnarray}  
$\omega_{0i}$ is the $i^{\rm th}$ electron frequency, $Z_j$ is the atomic number of $j^{\rm th}$ atom, while $f_i$ denotes the strength factor of an $i^{\rm th}$ polarizable electron in a given atom. Equation~(\ref{eq:6}) is obtained by neglecting the spontaneous emission (or classically known as the damping factor)~\cite{arul6}. Apparently, both $f(E_0,\xi)$ and $\alpha(\xi)$ are inversely proportional to $\xi$. However, IET is inapplicable for free-electron (due to $\xi = 0$) and Fermi liquid metals (because $\xi$ is an irrelevant constant), which imply IET is only useful for quantum systems with `relevant' $\xi$ such that their Hamiltonians can be written in the form of Eq.~(\ref{eq:2}).     

\section*{3. Time reversal symmetry}

We have stated earlier in the introduction that a time-independent internal electric current induces static `external' magnetic field, which actually violates TRS. Here, we will prove that TRS is indeed broken in the presence of `internal' electric current, and in the absence of applied magnetic field as correctly hypothesized by Messiah~\cite{albert}. Here, the `applied' magnetic field is defined to originate from a source external to the quantum system under investigation, whereas, the `external' magnetic field originates from a constant internal electric current within the observed quantum system. In order to construct the logical proof for this violation, we first need to expose the properties of time reversal operation properly and correctly, going beyond the expositions presented in Ref.~\cite{albert}. Of course, the mathematical arguments and definitions invoked herein must always be compatible with the physical observations. In the subsequent paragraphs, we will derive the necessary equations and definitions required for a proper time-reversal operation, which then can be used to tackle TRS head-on, and to develop our logical proof on TRS violation due to static internal electric current.

\subsubsection*{3.1. Time reversal operation for electrons}

Time reversal symmetry in quantum mechanical systems deal with the dynamical state represented by the time-dependent wavefunction. The energy eigenvalue, $E$ obtained from the Schr${\rm \ddot{o}}$dinger equation (see Eq.~(\ref{eq:2}) or~(\ref{eq:3})) is unique because the wavefunction, $\Psi(\textbf{r},t)$ is also unique such that two different wavefunctions, $\Psi(\textbf{r},t)$ and $\Psi'(\textbf{r},t)$ are not allowed to be the solutions to the same Schr${\rm \ddot{o}}$dinger equation simultaneously with the same eigenvalue where $\Psi(\textbf{r},t) \neq \Psi'(\textbf{r},t)$. This is a physical requirement. However, TRS requires the time-reversed wavefunction, $\Psi(\textbf{r},t)^{\rm rev}$ to be another valid solution to the same Schr${\rm \ddot{o}}$dinger equation, with the same eigenvalue where $\Psi(\textbf{r},t)^{\rm rev} \neq \Psi(\textbf{r},t)$. This inequality ($\Psi(\textbf{r},t)^{\rm rev} \neq \Psi(\textbf{r},t)$) is in violation of the above physical requirement (a unique wavefunction corresponds (one-to-one) to a unique eigenvalue, or precisely $\Psi(\textbf{r},t) \neq \Psi'(\textbf{r},t)$ is not allowed). Therefore, the only option we have, which allows both $\Psi(\textbf{r},t)^{\rm rev}$ and $\Psi(\textbf{r},t)$ as distinct solutions to the same Schr${\rm \ddot{o}}$dinger equation (that gives the same eigenvalue), and satisfies the requirement, $\Psi(\textbf{r},t)^{\rm rev} \neq \Psi(\textbf{r},t)$ is to define $\Psi(\textbf{r},t)^{\rm rev} = \Psi^*(\textbf{r},-t)$. We cannot define $\Psi(\textbf{r},t)^{\rm rev} = \Psi(\textbf{r},-t)$ because by definition, $\Psi(\textbf{r},-t) = \Psi(\textbf{r},t)$. Having said that, we can now easily show that $\Psi^*(\textbf{r},-t)$ is also a solution to the same Schr${\rm \ddot{o}}$dinger equation by taking the complex conjugate (both sides) of Eq.~(\ref{eq:3}) (after replacing $t$ with $-t$) to arrive at, 
\begin {eqnarray}
{\rm i}\hbar\frac{\partial}{\partial t}\Psi^*(\textbf{r},-t) = H\Psi^*(\textbf{r},-t) = E\Psi^*(\textbf{r},-t). \label{eq:7}
\end {eqnarray}  
As a consequence of the above complex conjugation, we can define the time reversal operator, $\texttt{K}$ properly by writing $\Psi(\textbf{r},t)^{\rm rev}$ in the form, 
\begin {eqnarray}
\Psi(\textbf{r},t)^{\rm rev} = \texttt{K}\Psi(\textbf{r},-t)\texttt{K}^* = \Psi^*(\textbf{r},-t),\label{eq:8}
\end {eqnarray}  
which means that the time reversal operation, in this case, is a form of complex conjugation~\cite{albert} because the phase factor is time-dependent, and it is not a constant. This operation will leave the real numbers (position, $\textbf{r}$) and real operators ($Q$) alone ($\texttt{K}\textbf{r}Q\texttt{K}^{*} = \textbf{r}Q$). In contrast, if $\texttt{K}$ operates on complex numbers or operators, then $\texttt{K}$ changes the sign in the following way, $\texttt{K} {\rm i}\texttt{K}^{*} = -{\rm i}$. Since $\texttt{K}$ is a complex conjugation operator, one can readily surmise that $\texttt{K} = \texttt{K}^{*}$. 

For an electron however, $\texttt{K}$ is insufficient because an electron is a spin-$\frac{1}{2}$ particle such that we also need to invert the spin (or change its sign) where each spin component ($\textbf{S}_x$, $\textbf{S}_y$ and $\textbf{S}_z$) should be inverted. This means that, we need a new time reversal operator, $\texttt{K}_{\textbf{S}}$, to have the following essential property, 
\begin {eqnarray}  
\frac{\hbar}{2}\texttt{K}_{\textbf{S}}\sigma_{x,y,z}\texttt{K}_{\textbf{S}}^{\dagger} = -\frac{\hbar}{2}\sigma_{x,y,z},\label{eq:9}
\end {eqnarray}  
where $\textbf{S}_{x,y,z} = \frac{\hbar}{2}\sigma_{x,y,z}$, $\sigma_{x,y,z} = \sigma_{x}, \sigma_{y}, \sigma_{z}$, denote the Pauli matrices and $\texttt{K}_{\textbf{S}}^{\dagger}$ is an antiunitary Hermitian conjugate. This inversion (sign change) due to time reversal operation is mandatory because the spin is intrinsically undefined (the spin is not static). From Eq.~(\ref{eq:8}), $\texttt{K}$ is a complex conjugation operator, and therefore, it cannot invert the real spin components ($\sigma_{x}$ and $\sigma_{z}$), which explains why we need a new time-reversal operator for electrons that has to be defined in a different context. This new operator, $\texttt{K}_{\textbf{S}}$ has to be a 2$\times$2 matrix such that $\texttt{K}_{\textbf{S}}$ also needs to invert the real Pauli matrices, namely, $\sigma_x$ and $\sigma_z$. This new time-reversal operator for electrons has been defined by Messiah~\cite{albert},
\begin {eqnarray}  
\texttt{K}_{\textbf{S}} = -{\rm i}\sigma_y\texttt{K}, \label{eq:10} 
\end {eqnarray}  
using a phase-factor type rotation operator. However, we still need to define $\texttt{K}_{\textbf{S}}^2$ properly, such that i in Eq.~(\ref{eq:10}) is left intact because it has a precise physical meaning with respect to spin via the rotation operator. We also need to know $\texttt{K}_{\textbf{S}}^2$ so that we can use it to generalize and to understand the time reversal operation for even and odd number of electrons. In particular, we can readily define $\texttt{K}_{\textbf{S}}^2 = -{\rm i}\sigma_y\texttt{K}(\sigma_{x,y,z}\psi(\textbf{r},t)){\rm i}\sigma_y\texttt{K}^*$ where we also have defined an arbitrary spin($\sigma$)-wavefunction($\Psi$), $\Psi(\sigma,\textbf{r},t) = \sigma_{x,y,z}\psi(\textbf{r},t)$, which is a complex number. One can readily recall the following useful identities, $\sigma_y\sigma_x\sigma_y = -\sigma_x$, $\sigma_y\sigma_y\sigma_y = -\sigma_y$, $\sigma_y\sigma_z\sigma_y = -\sigma_z$, $\sigma_y\sigma_y = \sigma_z\sigma_z = \sigma_x\sigma_x = \textbf{\textsl{I}}_2$, $\det$($\textbf{\textsl{I}}_2$) = 1 and $\texttt{K}\texttt{K}^* = \texttt{K}^2 = 1$. For a single electron, $\texttt{K}_{\textbf{S}}^2 = -{\rm i}\sigma_y(-\sigma_{x,y,z})\psi^*(\textbf{r},-t)(-{\rm i})(-\sigma_y)\texttt{K}\texttt{K}^* = -1\sigma_{x,y,z}\psi^*(\textbf{r},-t)$. For two electrons, $\texttt{K}_{\textbf{S}}^2 = [-1\sigma_{1;x,y,z}\psi^*_1(\textbf{r},-t)][-1\sigma_{2;x,y,z}\psi^*_2(\textbf{r},-t)] = [\sigma_{1;x,y,z}\psi^*_1(\textbf{r},-t)][\sigma_{2;x,y,z}\psi^*_2(\textbf{r},-t)]$. Therefore, for $n$ number of electrons, 
\begin {eqnarray}  
&&\texttt{K}_{\textbf{S}}^2 = (-1)^n\big[\sigma_{1;x,y,z}\psi^*_1(\textbf{r},-t)\big]\big[\sigma_{2;x,y,z}\psi^*_2(\textbf{r},-t)\big]\cdots\big[\sigma_{n;x,y,z}\psi^*_n(\textbf{r},-t)\big]. \nonumber \\&& \label{eq:11} 
\end {eqnarray}  
For even $n$, we can rewrite Eq.~(\ref{eq:11}) (recall the conditions used to derive Eq.~(\ref{eq:7})), 
\begin {eqnarray}
[\sigma_{1;x,y,z}\psi^*_1(\textbf{r},-t)][\sigma_{2;x,y,z}\psi^*_2(\textbf{r},-t)] \cdots &\longrightarrow& [\psi^{\rm real}_1(\textbf{r},-t)][\psi^{\rm real}_2(\textbf{r},-t)] \cdots \nonumber \\&& = [\psi^{\rm real}_1(\textbf{r},t)][\psi^{\rm real}_2(\textbf{r},t)] \cdots, \nonumber \\&& \label{eq:12} 
\end {eqnarray}
where we have dropped the spin components (or the Pauli matrices) because real wavefunctions cannot be defined as $\Psi(\sigma,\textbf{r},t) = \sigma_{x,y,z}\psi(\textbf{r},t)$. This means that, a real wavefunction needs to transform in this way, $\texttt{K}_{\textbf{S}}\psi^{\rm real}_1(\textbf{r},t)\texttt{K}^{\dagger}_{\textbf{S}} = \psi^{\rm real}_1(\textbf{r},t)$. Moreover, each electron in a system with even number of electrons (see Eq.~(\ref{eq:12})) can be represented by a real wavefunction, $\psi^{\rm real}_1(\textbf{r},t), \cdots, \psi^{\rm real}_n(\textbf{r},t)$. Therefore, to satisfy TRS for $n = n_{\rm even}$, we can either use real $\psi^{\rm real}_n(\textbf{r},t)$ or complex wavefunctions, $\sigma_{n;x,y,z}\psi_n(\textbf{r},t)$. If we were to use complex wavefunctions, then their eigenvalues must be at least doubly degenerate (see the following paragraph). 

On the other hand, for odd number of electrons, $\texttt{K}_{\textbf{S}}\sigma_{1;x,y,z}\psi_1(\textbf{r},t)\texttt{K}^{\dagger}_{\textbf{S}} = -\sigma_{1;x,y,z}\psi^*_1(\textbf{r},-t)$ where $-\sigma_{1;x,y,z}\psi^*_1(\textbf{r},-t)$ is orthogonal to $\sigma_{1;x,y,z}\psi_1(\textbf{r},t)$ and consequently, the eigenvalue for $-\sigma_{1;x,y,z}\psi^*_1(\textbf{r},-t)$ and $\sigma_{1;x,y,z}\psi_1(\textbf{r},t)$ is (doubly) degenerate. Contrary to even number of electrons, odd $n$ strictly require complex wavefunctions to satisfy TRS, and for $n_{\rm odd} > 1$, the eigenvalue is at least two-fold degenerate (with even order) due to Kramers degeneracy. In summary, by definition TRS requires gapless (degenerate) energy levels. Add to that, a gapless electronic system with even number of electrons can be represented by a set of real or complex wavefunctions to satisfy TRS. On the other hand, the same system with odd number of electrons needs complex wavefunctions so as not to violate TRS.

\subsubsection*{3.2. Time reversal violation due to electric current}

From the preceding subsections, we have formally and properly shown that any form of electron excitation, from the valence to conduction band violates TRS such that degenerate energy levels are mandatory for TRS---because we also need to independently time-reverse the photon that is responsible for the said excitation. Here, we claim that TRS is violated if there is a static internal electric current within a gapless quantum system. This is not a trivial claim, which exists only as a hypothesis due to Messiah~\cite{albert}, and therefore, we need some logical arguments to understand it properly. We consider a one-dimensional system for convenience, and using an arbitrary wavefunction, $\Psi(x,-t)$ where $\Psi_{\rm rev}(x,t) = \Psi^*(x,-t)$ (from Eq.~(\ref{eq:8})), one can write the flow of probability in this form,
\begin {eqnarray}
\frac{\partial}{\partial t}|\Psi(x,-t)|^2 = \Psi^*(x,-t)\frac{\partial\Psi(x,-t)}{\partial t} + \frac{\partial\Psi^*(x,-t)}{\partial t}\Psi(x,-t) \neq 0,  \label{eq:13} 
\end {eqnarray}
where $\Psi^*(x,-t)$ is a complex conjugate of $\Psi(x,-t)$, and $-t$ is kept explicit for we are attempting to derive the time reversed probability current, $\textbf{J}_{\rm rev}$. We can now rewrite Eq.~(\ref{eq:3}) to obtain,
\begin {eqnarray}
\frac{\partial \Psi^*(x,-t)}{{\rm d}t} = \frac{{\rm i}\hbar}{2m}\bigg[\frac{\partial^2\Psi^*(x,-t)}{\partial x^2} - \frac{{\rm i}}{\hbar}V\Psi^*(x,-t)\bigg].  \label{eq:14} 
\end {eqnarray}
Taking the complex conjugate of Eq.~(\ref{eq:14}), and using Eq.~(\ref{eq:13}), we arrive at the time reversed probability current,
\begin {eqnarray}
\textbf{J}_{\rm rev} = -\frac{{\rm i}\hbar}{2m}\bigg[\frac{\partial \Psi(x,-t)}{\partial x}\Psi^*(x,-t) - \Psi(x,-t)\frac{\partial \Psi^*(x,-t)}{\partial x}\bigg]. \label{eq:15} 
\end {eqnarray}
Now, using Eq.~(\ref{eq:3}) directly, we can derive the forward probability current,
\begin {eqnarray}
\textbf{J}_{\rm fwd} = \frac{{\rm i}\hbar}{2m}\bigg[\Psi(x,t)\frac{\partial \Psi^*(x,t)}{\partial x} - \frac{\partial \Psi(x,t)}{\partial x}\Psi^*(x,t)\bigg], \label{eq:16} 
\end {eqnarray}
or alternatively, Eq.~(\ref{eq:16}) can also be obtained from $\texttt{K}_{\textbf{S}}\textbf{J}_{\rm rev}\texttt{K}^{\dagger}_{\textbf{S}}$ after replacing $-t$ with $t$ and $\texttt{K}_{\textbf{S}}(\partial \Psi^*(x,t)/\partial x)\texttt{K}^{\dagger}_{\textbf{S}} = \partial \texttt{K}_{\textbf{S}}\Psi^*(x,t)\texttt{K}^{\dagger}_{\textbf{S}}/\partial x$. Next, we invoke one of the Maxwell equations within matter in the presence of macroscopic static internal current density (current per unit length, $\textbf{i}$),
\begin {eqnarray}
\nabla \times \textbf{B}' = \mu_{\rm perm}\textbf{i}_{\rm fwd} \propto \mu_{\rm perm}\textbf{J}_{\rm fwd}, \label{eq:17} 
\end {eqnarray}
where $\mu_{\rm perm}$ is the permeability constant and the term, `static current' here of course implies $\partial \textbf{D}/\partial t = 0$ in which, $\textbf{D}$ is the usual time-dependent electric charge displacement. Equation~(\ref{eq:17}) for $\textbf{i}_{\rm rev}$ is given by,
\begin {eqnarray}
\nabla \times \textbf{B}'' \propto \mu_{\rm perm}\textbf{J}_{\rm rev}, \label{eq:18} 
\end {eqnarray}
where $\textbf{B}'$ is the induced external magnetic field due to $\textbf{i}_{\rm fwd}$, while $\textbf{B}''$ is induced by $\textbf{i}_{\rm rev}$ such that $\textbf{B}'$ is or is not equal to $\textbf{B}''$, and it does not matter. The point is, both $\textbf{B}'$ and $\textbf{B}''$ are independent to each other (they have independent sources, $\textbf{i}_{\rm fwd}$ and $\textbf{i}_{\rm rev}$), they are not zero ($\textbf{i}_{\rm fwd}$ and $\textbf{i}_{\rm rev}$ are not zero) and they are not `applied' magnetic fields ($\textbf{i}_{\rm fwd}$ and $\textbf{i}_{\rm rev}$ are the currents within the system). Therefore, when the current is time-reversed, we also need to independently reverse $\textbf{B}'$. The requirement to time-reverse $\textbf{B}'$ independently when the current ($\textbf{i}_{\rm fwd}$) is reversed strictly means that TRS will be violated if we reverse only the current. In summary, TRS is guaranteed to be violated even in the presence of static (constant) electric current, which also induces the static `external' magnetic field. Here, we do not require any applied magnetic field to violate TRS, as correctly hypothesized by Messiah~\cite{albert}.

\textit{Warning}: One should be careful here to understand the points stated above on TRS based on Eqs.~(\ref{eq:17}) and~(\ref{eq:18}). For example, we cannot time-reverse $\textbf{B}'$ in order to reverse the current, $\textbf{i}_{\rm fwd}$ because the time reversed $\textbf{B}'$ will not induce its own current, in agreement with our analyses above. If it induces its own current, $\textbf{i}_{\rm rev}$, then $\textbf{B}' = \textbf{B}''$, which means $\textbf{B}''$ exactly cancels $\textbf{B}'$, and this cancellation symmetry has got nothing to do with our time reversal operation. This is why physical time reversal operation is entirely different from the philosophical one where the latter also implies going back in time by deleting the future. In particular, if we reverse $\textbf{B}'$ by deleting (or canceling) $\textbf{B}'$, then $\textbf{i}_{\rm fwd}$ is also deleted simultaneously, without the need to delete $\textbf{i}_{\rm fwd}$ independently. But the time reversal operation by erasing the future is never allowed in our universe (due to the second law of thermodynamics), and this difference (between the physical and philosophical time reversal operations) is the root cause for the confusion in TRS.    

\subsubsection*{3.3. Relevant and irrelevant energy-level spacing}

`Relevant' energy level spacing means a particular physical property is directly influenced by $\xi$, while an `irrelevant' $\xi$ and $\xi = 0$ have no influence at all. Examples of trivially relevant energy level spacings~\cite{arul7} are the atomic energy level spacings~\cite{web,web2}, band gaps in band insulators and semiconductors~\cite{ash}, Mott-Hubbard gaps in Mott insulators~\cite{mott}, and the molecular gaps in molecules~\cite{ira} (between the highest occupied molecular orbital (HOMO) and the lowest unoccupied molecular orbital (LUMO)). All Fermi liquid metals~\cite{ash} (including quantum Hall metals~\cite{klit,tsui}) require irrelevant $\xi$ ($\xi_{\rm irr} \neq 0$)~\cite{and1}, while $\xi = 0$ is for free electrons (Fermi gas)~\cite{arul7}. However, there is a special situation that can give rise to a nontrivially relevant $\xi$, which is applicable to strange metals in cuprates~\cite{arul7}, and other strongly correlated metals where $\xi$ is `special' because it exists within a degenerate electronic system, and its origin is due to different wavefunctions~\cite{arul8}. In particular, the special energy level spacing theorem reads~\cite{arul7,arul8}---an electron needs energy supply to occupy another degenerate energy level because the wavefunction representing this particular electron needs to be transformed (beyond the phase factor) to occupy the new energy level. 

In summary, from our previous analyses on TRS, it should be clear now that TRS will be violated in the presence of static electric current regardless of whether $\xi = 0$, or $\xi$ is an irrelevant constant, or $\xi$ is a finite relevant constant. This also means that any violation of TRS does not necessarily imply a given system is gapped because internal electric current has been shown to violate TRS even in gapless systems. Apart from that, even though TRS is known to be responsible for Kramers degeneracy, TI can form non-Kramers degenerate surface states due to many other complicated interactions as pointed out in the introduction, and in Refs.~\cite{arul7,arul8}.       

\section*{4. Results and discussion} 

We show here that the metallic property in gapless TI, can be associated to the above special (nontrivially relevant) energy level spacing ($\xi^{\rm non}_{\rm triv}$). Usually, a given compound is defined to be metallic if its resistivity ($\rho(T)$) gets smaller with lowering the temperature $T$. Both $\rho(T, x)$ and the carrier-type transition ($\texttt{n}$- to $\texttt{p}$-type) have been measured by Jinsong Zhang \textit{et al}.~\cite{jins} in (Bi$_{1-x}$Sb$_x$)$_2$Te$_3$ TI. Apart from metallic resistivity, we will also explain why the bandgap of another TI, namely, Pb$_{1-x}$Sn$_x$Se changes with Sn substitution, which were first measured by Strauss~\cite{strau} and later confirmed by Dziawa \textit{et al.}~\cite{dzia} using ARPES measurements. The latter bandgap analyses are based on the trivially relevant $\xi_{\rm triv}$. Since the energy level spacings for both gapless and gapped systems are nonzero and relevant, we are therefore forced to identify them with $\xi^{\rm non}_{\rm triv}$ and $\xi_{\rm triv}$ for gapless and gapped systems, respectively.

\subsubsection*{4.1. Carrier density and electron-ion interaction} 

Changes in the magnitude of resistivity in TI or any other strongly correlated solid state systems are usually due to doping $x$, $T$ and $\textbf{B}$. In the absence of $\textbf{B}$, and for constant $T$, the microscopic physical parameters that will significantly influence the resistivity with respect to $x$ are the carrier density ($n(\xi)$) and the scattering amplitude, $|f(\theta)|$. Here~\cite{arul2,arul5},  
\begin {eqnarray}
n(\xi) = \texttt{C}(T)\exp{\big[\lambda(\xi - E_{\rm F}^0)\big]}, \label{eq:19} 
\end {eqnarray}
where $\texttt{C}(T)$ denotes a collection of fundamental constants, which actually depends on the dimensionality of the system and $\texttt{C}(T)$ also contains a $T$-dependent parameter. The resistivity,
\begin {eqnarray}
\rho(x) = \frac{m}{ne^2\tau} = \texttt{C}'\tau(x)^{-1}\exp{\big[\lambda(\xi - E_{\rm F}^0)\big]}, \label{eq:20} 
\end {eqnarray}
where we have suppressed the $T$ dependence because $\rho(x)$ is for constant $T$. Next, we need to find how the scattering rate, $\tau(x)^{-1}$ vary with doping. For free-electron and Fermi liquid metals, $\tau(x)^{-1} \longrightarrow \tau^{-1}$ by definition~\cite{arul7} because the electrons are completely independent of the types of ions such that they interact with neutral phonons, not with the positively charged ions. This means that, $\tau^{-1}$ does not systematically change with the type of atoms in Fermi liquid~\cite{arul7}. In this case, the scattering rates are mostly contributed by the well-known electron-electron (e:e) and electron-phonon (e:ph) type collisions. On the other hand, in strongly correlated metals, including metallic TI, we have the usual contribution from $\tau_{\rm e:e}^{-1}$ and another dominant contribution from the electron-ion (e:ion) scattering rate, $\tau(x)_{\rm e:ion}^{-1}$. We show here that $\tau(x)_{\rm e:ion}^{-1}$ is proportional to the scattering amplitude, $|f(\theta)|$.

To understand why $\tau(x)_{\rm e:ion}^{-1} \propto |f(\theta)|$, we use the Born approximation~\cite{davi} and the renormalized screened Coulomb potential~\cite{arul5},
\begin {eqnarray}
V_{\rm e:ion} = \frac{e^2}{4\pi\epsilon_0r}\exp{\big[-\mu_{\rm IET}r e^{-(1/2)\lambda\xi}\big]}, \label{eq:21} 
\end {eqnarray}
to obtain the scattering amplitude,
\begin {eqnarray}
|f(\theta)| \cong \frac{2me^2}{4\pi\epsilon_0\hbar^2(\mu^2_{\rm IET}e^{-\lambda\xi} + \kappa^2)} , \label{eq:22} 
\end {eqnarray}
where $\mu_{\rm IET}$ is the constant of proportionality, and as required, Eq.~(\ref{eq:22}) reduces to the Rutherford scattering amplitude (for two-point charges) when $\xi \rightarrow \infty$~\cite{arul4}, and it also reduces to the Yukawa scattering amplitude if $\xi \rightarrow 0$~\cite{arul4}. Moreover, $\kappa = |\textbf{k}' - \textbf{k}|$ depicts the changes in the wavevector such that $\textbf{k}'$ and $\textbf{k}$ point in the incident and scattered directions, respectively, and during the process, the momentum transfer is given by $\hbar(\textbf{k} - \textbf{k}')$. It is obvious from Eq.~(\ref{eq:22}) that $\theta$ is the angle due to scattering between the directions, $\textbf{k}$ and $\textbf{k}'$. In addition, Eq.~(\ref{eq:22}) leads us straight to $|f(\theta)| \propto \xi$ that implies large scattering amplitude is obtained for an ion (scattering center) with large $\xi$, which in turn justifies the correctness of this relation, $\tau(x)_{\rm e:ion}^{-1} \propto |f(\theta)|$. As a consequence, the approximate resistivity can be written in the form (using Eq.~(\ref{eq:20})),
\begin {eqnarray}
\rho(x) \propto \frac{\texttt{C}''\exp{\big[\lambda(\xi - E_{\rm F}^0)\big]}}{\mu^2_{\rm IET}e^{-\lambda\xi} + \kappa^2}, \label{eq:23} 
\end {eqnarray}
where $\texttt{C}''$ is another collection of fundamental constants, which includes $\texttt{C}$ and $\texttt{C}'$. Firstly, even though Eq.~(\ref{eq:23}) is an approximate one, but it exactly captures the sought-after effect of changing carrier density and e:ion scattering strength on $\rho(x)$, and secondly, the equation also shows that $n(\xi)$ and $\tau^{-1}(x)$ do not compete with each other with opposite effects as a result of doping or changing chemical composition. This second point is important such that $\rho(x)$ (for constant $T$) is guaranteed to decrease if one substitutes a chemical element with another one that has a smaller $\xi$ because small $\xi$ leads to large $n(\xi)$ and $\tau(x)$ (see Eq.~(\ref{eq:20})). Thus far, we have ignored the $T$-dependence even though we know both $\xi(T)$ and $\tau_{\rm e:ion}(T)$ do exist simply because their explicit forms are unknown due to their complex $T$-dependences. Logically, $\tau_{\rm e:ion}$ has to be $T$-dependent, while $\xi$ has been experimentally proven to be $T$-dependent by Dionicio~\cite{dio} using the results of Fukuda \textit{et al.}~\cite{fuku}. In particular, Dionicio showed that the valence state of a given multivalent chemical element is $T$-dependent. But anyway, Eq.~(\ref{eq:23}) is exact with respect to $\xi$-dependence, and therefore, it is sufficient for our analyses, and to show that indeed non-Kramers degeneracy (due to relevant $\xi^{\rm non}_{\rm triv} \neq 0$) is responsible for the metallic surface states in TI.      

\subsubsection*{4.2. Doping-dependent resistivity and carrier-type transition}

We are basically done deriving all the equations needed to explain the measured resistivity data and the carrier-type transition in Pb$_{1-x}$Sn$_x$Se and (Bi$_{1-x}$Sb$_x$)$_2$Te$_3$ TI. Dziawa \textit{et al.}~\cite{dzia} have studied the narrow bandgap semiconductor, Pb$_{1-x}$Sn$_x$Se where a topological phase transition is observed for $x = 0.23$ when $T$ is reduced from 300K to 77K. In particular, Pb$_{0.77}$Sn$_{0.23}$Se is a gapped semiconductor for $T = 300$K, and it becomes a gapless TI for $T = 77$K due to band inversion observed indirectly via ARPES. Moreover, for a given $T$ (300K or 195K or 77K), Sn substitutional doping (increasing $x$ from 0 to about 0.3) systematically reduces the bandgap of Pb$_{1-x}$Sn$_x$Se semiconductor from about 0.3 to approximately 0.05 eV (see Fig.~1 in Ref.~\cite{dzia}). This observation is easily understood within IET by noting that $\xi$ now represents the bandgap, $\xi_{\rm Sn^{2+}} < \xi_{\rm Pb^{2+}}$ (see Table~1), and from Eqs.~(\ref{eq:19}) and~(\ref{eq:23}), $\rho(x)$ is predicted to decrease with increasing Sn content, provided that the valence states of the chemical elements (Pb and Sn) do not change significantly due to defects. If the valence states do change, then the analyses become tedious, which have been addressed elsewhere for other strongly correlated materials and oxides~\cite{arul7,arul9,arul10} where topological insulators are no exception (see below).

In the absence of defects, the changes on $\rho(x)$ for Pb$_{1-x}$Sn$_x$Se have been exposed in a straightforward manner within IET. However, the (Bi$_{1-x}$Sb$_x$)$_2$Te$_3$ material is reported to have a complicated doping-dependent resistivity by Jinsong Zhang \textit{et al.}~\cite{jins}. For example, the measured $\rho(x)$ first increases with $x$, for up to $x = 0.94$, and then it reduces until $x = 1$ (see Fig.~4 in Ref.~\cite{jins}). With increasing Sb content ($0 \leq x \leq 0.94$), the system also become relatively more insulating. Further Sb substitutional doping ($0.94 < x \leq 1$) leads to a weaker insulating behavior such that Sb$_2$Te$_3$ is a semimetal with holes as the dominant charge carriers. In contrast, Bi$_2$Te$_3$ is an electron-dominant metal. The above carrier-type transition from an $\texttt{n}$-type insulator to a $\texttt{p}$-type semimetal are of course due to defects where the valence states for Sb$^{3+}$ and Te$^{2+}$ are not constants with doping (for $x > 0.94$). 

Similar to Si = [Si$^{4+}$][Si$^{4+}$] semiconductor, (Bi$_{1-x}$Sb$_x$)$_2$Te$_3$ is also predominantly covalent bonded and therefore, we can also write it in the form, [(Bi$_{1-x}$Sb$_x$)$_2$]$^{6+}$[Te$_3$]$^{6+}$ = [(Bi$^{3+}_{1-x}$Sb$^{3+}_x$)$_2$][Te$^{2+}_3$]. Using Eqs.~(\ref{eq:19}) and~(\ref{eq:23}), and the fact that $\xi_{\rm Bi^{3+}} < \xi_{\rm Sb^{3+}}$, one can deduce that $\rho(x)$ should increase with increasing Sb content, which has been observed for $0 \leq x \leq 0.94$. Next, to understand the carrier-type transition in this class of material, it is sufficient for us to focus on these two extreme cases, namely, for $x = 0$ ($\texttt{n}$-type; Bi$^{3+}_{2}$Te$^{2+}_3$) and for $x = 1$ ($\texttt{p}$-type; Sb$^{3+}_2$Te$^{2+}_3$). We will exploit the carrier-type transition theorem developed in Ref.~\cite{arul10}, which will also lead us to understand why $\rho(x)$ have decreased for $x > 0.94$.              

The carrier-type transition theorem states that $\texttt{p}$-type materials with relevant $\xi$ should satisfy this condition~\cite{arul10}, 
\begin {eqnarray} 
\xi_{\rm acceptor}^{a+} < \xi_{\rm host}^{h+},~~~ a < h ~~~{\rm and} ~~~x_{\rm acceptor} < y_{\rm host}, \label{eq:24} 
\end {eqnarray} 
where $a$ and $h$ denote the acceptor and host valence states, respectively, while $x_{\rm acceptor}$ and $y_{\rm host}$ are the respective concentrations of the acceptor and host chemical elements. The acceptor accepts holes from the host such that the electrons from the acceptor are more easily polarizable than that of the host following Eq.~(\ref{eq:6}). For example, for [(Bi$^{3+}_{1-x}$Sb$^{3+}_x$)$_2$][Te$^{2+}_3$], the subscripts, $2(1-x)$ and 3 are the concentrations for the host chemical elements, Bi and Te for $x < 0.5$. Now, Bi$^{3+}_{2}$Te$^{2+}_3$ is clearly an $\texttt{n}$-type system because it does not fulfill the above condition where Te$_3$ is both the host and acceptor, which is not allowed from Eq.~(\ref{eq:24}), and if this is the case, then Eq.~(\ref{eq:6}) does not allow the creation of holes~\cite{arul10}. In other words, the inequality, $\xi_{\rm acceptor}^{a+} > \xi_{\rm host}^{h+}$ does not exist because $\xi_{\rm acceptor; Te}^{2+} = \xi_{\rm host; Te}^{2+}$ where the latter equality violates the condition stated in Eq.~(\ref{eq:24}), and therefore, Eq.~(\ref{eq:6}) cannot be used to create holes. Given this background, Sb$^{3+}_2$Te$^{2+}_3$ is also predicted to be an $\texttt{n}$-type material. 

However, Sb$^{3+}_2$Te$^{2+}_3$ can be made to be a $\texttt{p}$-type material if we introduce some small amount of defects (with $x_{\rm new} \ll 2$ and $y \ll 3$) to give rise to [Sb$^{3+}_{2-x_{\rm new}}$Sb$^{a'+}_{x_{\rm new}}$][Te$^{2+}_{3-y}$Te$^{a''+}_{y}$]. Here, if $x_{\rm new}$ exists, then $y$ should also exist to balance the defects introduced by $x_{\rm new}$ so as to maintain a proper coordination number, regardless of the crystal growth conditions because the distribution of these defects need not be homogeneous at all. We now evaluate [Sb$^{3+}_{2-x_{\rm new}}$Sb$^{a'+}_{x_{\rm new}}$] and [Te$^{2+}_{3-y}$Te$^{a''+}_{y}$] separately. Apparently, Sb$^{3+}_{2-x_{\rm new}}$ is the host, and Sb$^{a'+}_{x_{\rm new}}$ is the acceptor where $\xi_{\rm Sb}^{3+} > \xi_{\rm Sb}^{a'+}$ if $a' < 3+$. Similarly, for [Te$^{2+}_{3-y}$Te$^{a''+}_{y}$], Te$^{2+}_{3-y}$ is the host, while Te$^{a''+}_{y}$ is the acceptor such that $\xi_{\rm Te}^{2+} > \xi_{\rm Te}^{a''+}$ if $a'' < 2+$. These inequalities satisfy Eq.~(\ref{eq:24}), and consequently, Sb$^{3+}_{2-x_{\rm new}}$Sb$^{a'+}_{x_{\rm new}}$Te$^{2+}_{3-y}$Te$^{a''+}_{y}$ is a $\texttt{p}$-type TI in the presence of defects. 

Moreover, $\xi_{\rm Sb}^{3+} > \xi_{\rm Sb}^{a'+}$ and $\xi_{\rm Te}^{2+} > \xi_{\rm Te}^{a''+}$ give rise to decreasing $\rho(x)$ for $x > 0.94$, which is in agreement with the observed data (see Fig.~4 in Ref.~\cite{jins}). The resistivity should decrease for $x > 0.94$ because the carrier-type transition to $\texttt{p}$-type is caused by the increasing defect density ($x_{\rm new}$ and $y$), namely, Sb$^{a'+}$ and Te$^{a''+}$ for $x > 0.94$, which give rise to increasing number of holes due to these inequalities, $\xi_{\rm Sb}^{3+} > \xi_{\rm Sb}^{a'+}$ and $\xi_{\rm Te}^{2+} > \xi_{\rm Te}^{a''+}$. Using the above analyses, one can also theoretically obtain a $\texttt{p}$-type Bi$^{3+}_{2}$Te$^{2+}_3$ with appropriate defects, Bi$^{a'+}$ and Te$^{a''+}$ where $a' < 3+$ and $a'' < 2+$. The above defect densities can be determined experimentally by using the chemical method to measure the valence states of various chemical elements that was first carried out by Mahendiran \textit{et al.}~\cite{mahen}. 

In summary, both Sb$^{3+}_{2}$Te$^{2+}_3$ and Bi$^{3+}_{2}$Te$^{2+}_3$ can be synthesized as $\texttt{p}$-type materials for $x_{\rm new} \neq 0$ and $y = 0$ or $x_{\rm new} = 0$ and $y \neq 0$, without requiring both $x_{\rm new} \neq 0$ and $y \neq 0$ as discussed above. We also have used the so-called relevant $\xi$ to explain the transport properties (in (Bi$_{1-x}$Sb$_x$)$_2$Te$_3$ and Pb$_{1-x}$Sn$_x$Se TI) and the $\texttt{n}$- to $\texttt{p}$-type carrier-type transition in (Bi$_{1-x}$Sb$_x$)$_2$Te$_3$, which unequivocally proves that the Kramers degeneracy is not responsible for the metallic surface states, at least in the above stated TI.           

\section*{5. Additional notes}

Using the proven TRS violation result, and the notion of energy-level spacing within IET, we have justified that these topological insulators, namely, (Bi$_{1-x}$Sb$_x$)$_2$Te$_3$ and Pb$_{1-x}$Sn$_x$Se, in the presence of surface electric current necessarily violate TRS. Even in the absence of IET, TRS is guaranteed to be violated for any solid state system (gapless or gapped) in the presence of electric current. Furthermore, within IET, we have shown that the degeneracy in topological insulators is not of the Kramers-type. To prove this, we derived the temperature-independent resistivity and scattering-rate equations as functions of doping ($x$). The experimental results obtained from the above materials ((Bi$_{1-x}$Sb$_x$)$_2$Te$_3$ and Pb$_{1-x}$Sn$_x$Se) follow exactly as predicted by these equations (Eqs.~(\ref{eq:23}) and~(\ref{eq:24})). This means that, the energy levels are crossed (in gapless systems) in such a way that there are nonzero wavefunction-induced energy gaps at the crossing points. This `special' gap exists due to different orthogonalized wavefunctions.

To reinforce the above non-Kramers degeneracy in topological insulators, we also provide precise explanation on the physics of $\texttt{n}$- to $\texttt{p}$-type carrier-type transition in (Bi$_{1-x}$Sb$_x$)$_2$Te$_3$. As a matter of fact, the existence of non-Kramers degeneracy in any system, correlated or not, is not surprising because complicated interactions and/or their interplay indeed have given rise to crossed energy levels. For example, in Fermi metals, metallic heavy fermions, conventional superconductors, strange metals (cuprates and other strongly correlated metals), metallic ferromagnets (manganites), quantum Hall metals (2-dimensional Fermi gas) and semi-metals.   

Our proof of TRS violation in any condensed matter system in the presence of electric current, directly challenges the analysis carried out by K${\rm \ddot{o}}$nig \textit{et al}.~\cite{konig} and Bernevig, Hughes and Zhang~\cite{bern2}. They~\cite{konig,bern2} claimed that the spin-down and spin-up (spin polarized) edge currents do not violate TRS due to Kramers degeneracy. We have proven that, with or without Kramers degeneracy, TRS will be violated in the presence of spinful electric current. Secondly, the degeneracy in topological insulators is not of the Kramers-type, which are unambiguously supported by the doping-dependent experimental results.   

\section*{6. Conclusions}

We have theoretically shown that the time reversal symmetry can be violated for all materials with non-zero energy level spacing, even in the presence of degenerate energy levels. In this case, the degeneracy is not of the Kramers-type due to the existence of nontrivially relevant energy level spacing, $\xi$. To prove the existence of non-Kramers degeneracy in topological insulators, namely, in (Bi$_{1-x}$Sb$_x$)$_2$Te$_3$ and Pb$_{1-x}$Sn$_x$Se, we have exploited the ionization energy theory to explain their transport properties and the $\texttt{n}$- to $\texttt{p}$-type carrier-type transition. In doing so, we have unambiguously shown that the nontrivially relevant $\xi$ played a pivotal role in the metallic states of topological insulators where Kramers degeneracy is not responsible for gapless system with nonzero and relevant $\xi$. In particular, we have explained the systematic changes to the band gap and the resistivity in Pb$_{1-x}$Sn$_x$Se TI as a result of $\xi$, which changes with different chemical compositions (due to changing $x$). In addition, $\xi$ is also found to be responsible for the $\texttt{n}$- to $\texttt{p}$-type carrier-type transition and doping-dependent (temperature-independent) resistivity in (Bi$_{1-x}$Sb$_x$)$_2$Te$_3$ TI. These physical phenomena were captured entirely by the nontrivially relevant energy level spacing even in the gapless metallic states of topological insulators. 

One can readily extend the analyses performed herein to other doped topological insulators where compatible dopants can be selected from the periodic table of chemical elements, and after estimating their most probable valence states, one can then calculate their average ionization energies to predict and explain the transport phenomena, even in the presence of defects.         

\section*{Acknowledgments}

I am grateful to the financiers, Sebastiammal Savarimuthu, Arulsamy Innasimuthu, Amelia Das Anthony, Malcolm Anandraj and Kingston Kisshenraj for their kind support and hospitality between Aug 2011 and Aug 2013. I am also grateful to Marco Fronzi (Osaka City University) for his unconditional help in providing most of the references.

\newpage
~\\ \textbf{Table~1}: Averaged atomic ionization energies ($\xi$) for selected atoms are ordered with increasing atomic number $Z$. The averaging follows Eq.~(\ref{eq:4}) and we used the unit kJmol$^{-1}$ for numerical convenience.

\newpage

Table~1
\begin{table}[ht]
\begin{tabular}{l c c c} 
\hline\hline 
\multicolumn{1}{l}{Element}          &    Atomic number    &  Valence         & $\xi$   \\  
\multicolumn{1}{l}{}                 &   $Z$               &  state           &(kJmol$^{-1}$)\\  
\hline 
Se                                   &  34   					    &  1+               & 941 \\ 
Se                                   &  	   	  			    &  2+               & 1493  \\ 
Se                                   &                    &  4+               & 2476 \\ 
\\ 
Sn                                   &  50	   	  			  &  1+               & 709 \\ 
Sn                                   &  	   	  			    &  2+               & 1061 \\
Sn                                   &  	   	  			    &  4+               & 1688 \\  
\\ 
Sb                                   &  51	   	  			  &  1+               & 834 \\ 
Sb                                   &  	   	  			    &  3+               & 1623 \\ 
\\
Te                                   &  52	   	  			  &  1+               & 869 \\ 
Te                                   &  	   	  			    &  2+               & 1330\\ 
Te                                   &  	   	  			    &  4+               & 2242 \\ 
\\ 
Pb                                   &  82	   	  			  &  1+               & 716 \\
Pb                                   &  	   	  			    &  2+               & 1084 \\
Pb                                   &  	   	  			    &  4+               & 2333 \\ 
\\ 
Bi                                   &  83	   	  			  &  1+               & 703 \\
Bi                                   &  	   	  			    &  3+               & 1593 \\
\hline  
\end{tabular}
\label{Table:I} 
\end{table}

\end{document}